\documentclass[preprint,aps,amsmath]{revtex4}
\usepackage{graphicx}
\usepackage{bm}

\begin{document}
\preprint{}
\title{Drift wave turbulence in a dense semiclassical magnetoplasma}
\author{A. Kendl}
\affiliation{\small Institute for Ion Physics and Applied Physics, 
University of Innsbruck, A-6020 Innsbruck, Austria}
\author{P. K. Shukla}
\affiliation{\small 1) RUB International Chair, International Centre for Advanced
  Studies in Physical Sciences, Faculty of Physics \& Astronomy, Ruhr
  University Bochum, D-44780 Bochum, Germany\\
2) Department of Mechanical and Aerospace Engineering, University
  of California San Diego, La Jolla, CA 92093, USA \vspace{2cm}}

\begin{abstract}
\bigskip
A semiclassical nonlinear collisional drift wave model for dense magnetized plasmas is
developed and solved numerically. The effects of fluid electron density fluctuations
associated with quantum statistical pressure and quantum Bohm force are
included, and their influences on the collisional drift wave instability and
the resulting fully developed nanoscale drift wave turbulence are discussed. 
It is found that the quantum effects increase the growth rate of the collisional 
drift wave instability, and introduce a finite de Broglie length screening 
on the drift wave turbulent density perturbations. The relevance 
to nanoscale turbulence in nonuniform dense magnetoplasmas is discussed. 

\vspace{4cm}

{\sl This is the preprint version of a manuscript submitted to Physics Letters A (2011).}
\end{abstract}
\maketitle

Recently, there has been considerable interest \cite{r1,r2,r3,r4} in the
investigation of collective dynamical processes in dense quantum plasmas, which are
ubiquitous in astrophysics (e.g. white dwarf  stars), in planetary systems
(e.g. Jupiter and giant planets around extraterrestrial stars), in
plasma-assisted nanotechnology (e.g. quantum diodes, quantum free-electron
lasers, metallic thin films and nanostructures, nanowires, nanoplasmonics,
etc.), and in highly compressed plasmas for inertial confinement fusion by
extremely powerful laser and charged particle beams.  

In quantum plasmas, the dynamics of degenerate electrons is governed by quantum hydrodynamic (QHD) 
equations \cite{r1} in which the electron momentum equation has novel forces associated 
with the electron pressure involving the Fermi-Dirac distribution function, and the 
quantum Bohm force arising from the overlapping of electron wavefunctions due 
to Heisenberg's uncertainty principle. Furthermore, in a dense plasma non-degenerate ions are in 
a strongly coupled state since the ratio between the Coulomb ion interaction energy and 
the ion kinetic energy is much larger than unity. Thus, kinematic ion viscosity and 
viscoelastic ion relaxation for ion correlations have to be included in the description of the 
ion motion. It then turns out that collective effects \cite{r1,r2,r3,r4} in quantum plasmas arise 
at nanoscales due to relatively high plasma number density and the low electron and ion
temperatures when compared with the classical plasma. We thus have new quantum regimes, the physics 
of which is quite interesting and appealing for practical applications, as mentioned above.
    
The excitation of the electrostatic (ES) drift waves (DWs) \cite{r5,r6} is a universal feature of 
the classical nonuniform magnetoplasma. Classical ES DW turbulence has been mostly investigated by 
using the two-fluid theory and simulations \cite{Horton,Scott}, where
non-degenerate electron and ion fluids  
in the presence of the low-frequency (in comparison with the ion
gyrofrequency) drift waves are dynamically  
coupled via self-consistent ES fields and finite density fluctuations. It is widely thought that 
DWs are responsible for the cross-field transport of plasma particles in magnetic 
fusion devices \cite{Horton}.  The two-fluid quantum magnetohydrodynamic (Q-MHD) equations for 
a uniform quantum magnetoplasma, by including electron-$1/2$ spin effect, have been developed 
by Brodin \cite{Brodin07}, by assuming that non-degenerate ions are uncorrelated.  Furthermore, 
the linear properties of the ES DWS in a nonuniform quantum magnetoplasma have been studied 
by several authors \cite{Shokri99,Ali07}. 

In this Letter, we develop nonlinear equations for the low-frequency ES DWs in a nonuniform 
quantum magnetoplasma that is collisional. The governing mode-coupling equations are then numerically 
solved to depict the novel features of fully developed ES DW turbulence at nanoscales in 
a driven (due to the combined action of the density gradient and electron-ion collisions), 
nonuniform Fermi magnetoplasma. 

Let us consider a nonuniform quantum magnetoplasma in an external magnetic field ${\bf B}$.
A background plasma density gradient $\nabla n_0(x)$ is perpendicular to the
magnetic field direction ${\bf b} = {\bf B}/B$ along $z$ in a local Cartesian
coordinate system.

The governing nonlinear fluid equations for electron with density $n_e({\bf x},t)$ and
velocity ${\bf v}({\bf x},t)$, and ions with density $n_i({\bf x},t)$ and
velocity ${\bf u}({\bf x},t)$ are the continuity equations  
\begin{eqnarray}
 (\partial_t + {\bf v} \cdot \nabla ) n_e + n_e \nabla \cdot {\bf v} =  0, & &  \\
 (\partial_t + {\bf u} \cdot \nabla )  n_i + n_i \nabla \cdot {\bf u}  =  0, & &  
\label{continuity} 
\end{eqnarray}
and the momentum equations.
The inertialess momentum equation for degenerate electrons is
\begin{equation}
 {\bf F}_e  - {1 \over n_e} \nabla p_e +   {\bf F}_B  - m_e \nu_{ei}({\bf
   v}\!-\!{\bf  u}) = 0, 
\label{electron} 
\end{equation}
where ${\bf F}_e = e \left(\nabla \phi -{\bf v} \times {\bf B} /c\right)$
is the Lorentz force on electrons with charge $e$ in a dynamical electrostatic potential
$\phi({\bf x},t)$ and static magnetic field ${\bf B}$.

The pressure term for the non-relativistic degenerate electrons is expressed as
$p_e = [T_e +(5/3) T_F ] n_e$, where the thermal electron
Temperature $T_e$ is in the following assumed to be much smaller than the Fermi electron temperature 
$T_F = (\hbar^2/2m_e) (3\pi^2n_0)^{2/3}$.
The Fermi energy for electrons with mass $m_e$ is $m_e v_F^2$ where 
$v_F= \sqrt{T_F/m_e}$ is the Fermi velocity at equilibrium density $n_0$. 

The effect of quantum hydrodynamical diffraction on electrons is included by the Bohm force 
${\bf F}_{B} = (\hbar^2/2 m_e) \nabla \left(\nabla^2 \ln \sqrt{n_e} \right)$. 
Electrons are coupled to ions via the collision frequency $\nu_{ei}$
representing the momentum exchange, with  $m_i n_i \nu_{ie} = m_e n_e \nu_{ei}$.

The momentum equation for inertial ions is 
\begin{equation}
n_i m_i (\partial_t + {\bf u} \cdot \nabla) {\bf  u}  +  n_i {\bf F}_i +
\nabla p_{i} + n_i m_i \nu_{ie}({\bf u}\!-\!{\bf v})  =  {\bf R}
\label{ion}
\end{equation}
where the viscous term for strongly coupled ions is determined by 
\begin{equation}
(1+ \tau_m \partial_t) {\bf R}  = \eta \nabla \! \cdot \! \nabla {\bf u} + (\xi
 + \eta/3) \nabla (\nabla \! \cdot \! {\bf u}) 
\end{equation}
with viscoelastic relaxation time $\tau_m$, and $\eta$ and $\xi$ are the
longitudinal and bulk viscosities \cite{r7,r8}, respectively.  

The Lorentz force on ions with charge state $Z$ is 
${\bf F}_i =  Z e \left(\nabla \phi - {\bf u} \times {\bf B} /c \right)$.

The ion pressure is $p_i = \gamma_i n_i (\mu_i T_i +
T_*)$, where $\gamma_i$ is the ion adiabatic index.
The excess chemical potential 
$\mu_i \approx 1+ 0.33 U(\Gamma_i) +(\Gamma_i/9)\partial U(\Gamma_i)/\partial \Gamma_i$,
includes $U (\Gamma_i)$ as a measure of the excess internal ion energy, where $\Gamma_i=Z_i^2/a_i T_i$,
$a_i =(3/4\pi n_i)^{1/3}$ is the Wigner-Seitz ion radius, and $T_i$ the
ion temperature. 
$T_* =(N_n/3)(Z_i^2e^2/a_i) (1+\kappa)\exp(-\kappa)$ accounts for the strong
ion coupling effect \cite{r9}. $N_n$ is determined by the ion structures and
corresponds to nearest neighbors ion interactions if the ions are in a
crystalline state,  
$\kappa =a_i/\lambda_{TF}$, and $\lambda_{TF} =v_F/\omega_{pe}$ the Thomas-Fermi screening radius, 
with $\omega_{pe} =(4\pi n_e e^2/m_e)^{1/2}$ being the electron plasma
frequency. 

In the following the drift approximation is applied, assuming that all
dynamical frequencies $\omega \ll \omega_{ce}$ are low compared to the
electron gyro frequency $\omega_{ce} = eB/m_ec$.
The velocities are expressed as ${\bf v} = {\bf v}_\perp + {\bf b} v_{||}$.
The perpendicular component of eq.~\ref{electron} gives the electron fluid
drift velocities ${\bf v}_{\perp} =  {\bf v}_E + {\bf v}_{F \ast} + {\bf v}_{q}$,
where ${\bf v}_E = (c/B^2) \; {\bf B} \times \nabla \phi$ is the classical
$E\times B$ drift velocity, 
${\bf v}_{F \ast} = (c T_F /e n n_0^2 B^2) \; {\bf B} \times \nabla n^3$
is the electron Fermi drift velocity,
and ${\bf v}_{q} = (c \hbar^2 / 2 e B^2 m_e n) \; {\bf B} \times \nabla_{\perp}
\left(\nabla^2 \ln \sqrt{n} \right)$ is the quantum diffraction drift velocity.  
Here we have assumed that $\nu_{ei} \ll \omega_{ce}$.
Quasi-neutrality requires $\tilde n_e \approx \tilde n_i \equiv n$.
The electron continuity equation in drift approximation is written as 
\begin{equation}
\left(\partial_t  + {\bf v}_E \cdot \nabla \right) n + n \nabla \cdot {\bf v}_E
- \nabla \cdot (n {\bf v}_{F \ast}) 
- \nabla \cdot (n  {\bf v}_{q})  - \nabla_{||} (n v_{z}) = 0. 
\label{e-dens1}
\end{equation}

In addition to the dominant convecting $E \times B$ velocity,
the drift velocities acquire a finite divergence in an inhomogeneous magnetic
field, which enters like a compressibility into the continuity equations by
abbreviating the curvature operator as 
\begin{equation}
\kappa (f)  =  -  c \nabla \times \left( { {\bf B} / B^2 } \right) \cdot \nabla f  
 =  - c \nabla \times \left[({\bf B} \times \nabla f)/B^2\right].  
\end{equation}
We obtain the divergence of  
the drift velocities as $ \nabla \cdot  {\bf v}_E = - \kappa (\phi)$, 
$ \nabla \cdot (n {\bf v}_{F \ast}) = (T_{F}/e) \kappa(n)$, 
and $\nabla \cdot (n  {\bf v}_{q}) =  - (\hbar^2 \kappa /4 m_e e) \nabla^2 n$.
The density $n = n_0(x) + \tilde{n}(x,y,t)$ can be now split into the background and
fluctuating components, viz. $n_0 (x)$ and $\tilde n(x,y,t)$, respectively, where 
$\tilde{n}(x,y.t) \ll n_0(x)$.  

The background gradient advection term can then be expressed as
${\bf v}_E \cdot \nabla n_0  = - (c/L_nB) \partial_y \phi$, 
with $L_n^{-1} = |\partial_x \ln n_0|$.  
In the following we drop the tilde on the fluctuating density component for clarity. 
Equation (6) then becomes
\begin{equation}
D_t n  =  {c n_0 \over B L_n} \partial_y \phi - {1 \over e} \nabla_{||} j_{||}  
- {T_{F} \over e} \kappa (n) 
  + \kappa(\phi) - {\hbar^2 \over 4m_e e} \kappa \left(\nabla^2 n \right), 
\label{e-dens2}
\end{equation}
where $D_t =\partial_t + {\bf v}_E \cdot \nabla$. The parallel electron current 
density $j_{||}$ is obtained from the momentum equation as
\begin{equation}
j_{||} = {n_0 e^2 \over m_e \nu_{ei}} \nabla_{||} \phi - {e T_F\over m_e \nu_{ei}}
\nabla_{||} (1-\lambda_q^2 \nabla^2) n
\end{equation}
where the electron de Broglie length at the Fermi temperature is 
$\lambda_q = \hbar / \sqrt{4 m_e T_{F}}$. 
The ions are assumed two-dimensional with $v_{||} \gg u_{||} \equiv 0$.

For non-degenerate ions, we assume that $\tau_m \partial_{t} {\bf u}_\perp \ll {\bf u}_\perp$, where 
the perpendicular component of the ion fluid velocity in the drift approximation 
($\omega \ll \omega_{ci} =Z_ieB/m_i c$) is $ {\bf u}_{\perp} = {\bf v}_E + {\bf u}_{p}$.
The $E\times B$ drift velocity for ions ${\bf  u}_E = {\bf v}_E$ is identical to
those for electrons, and the ion polarization drift velocity is 
$ {\bf u}_p \approx - (c/B \omega_{ci}) D_t \nabla_\perp \phi$. 
Here the ion gyrofrequency $\omega_{ci}$ is assumed to be much larger than $\mu_{ie}$, 
and we have assumed that $|D_t^2| \gg [(\xi +4\eta/3)/n_0 m_i] \nabla_\perp^2$. 
Assuming $T_i \approx T_e \ll T_F$, we have also excluded the contribution of
the ion diamagnetic drift in ${\bf u}_\perp$.  The ion continuity equation in
drift approximation then yields  
\begin{equation}
D_t n = {c n_0 \over B L_n} \partial_y \phi 
+ \kappa(\phi) - {c n_0 \over B \omega_{ci}} D_t \nabla_\perp^2 \phi ,
\label{i-dens}
\end{equation}

Subtracting (\ref{e-dens2}) from (\ref{i-dens}) we obtain the modified ion vorticity equation 
\begin{equation}
0  =  D_t \nabla^2_{\perp} {e \phi \over T_F} - {B \omega_{ci} \over n_0 e c} \nabla_{||} j_{||} 
- {T_{F} B \omega_{ci} \over n_0 e c} \kappa (n) 
 -{\hbar^2 B \omega_{ci}\over 4e n_0 m_e c} \kappa \left(\nabla^2 n \right). 
\label{div-j}
\end{equation}

Eqs.~(\ref{e-dens2}) and (\ref{div-j}) constitute coupled nonlinear dynamical equations
for low-frequency perturbations of the density $n$ and electrostatic potential
$\phi$ in an inhomogeneous  magnetised semi-classical plasma.
The standard drift normalization is applied, with the Fermi temperature $T_F$ as reference.
Spatial scales $x/L_{\perp} \rightarrow x$ are in units of a perpendicular
reference length $L_{\perp}$, and times scale as $t c_s / L_{\perp}
\rightarrow t$ with $c_s = \sqrt{T_F/m_i}$.
Fluctuating components are normalized as $\delta_0^{-1} e \phi / T_F
\rightarrow \phi$ and $\delta_0^{-1} n/n_0 \rightarrow n$,
where $\delta_0 = \rho / L_{\perp}$ with Fermi drift scale $\rho = \sqrt{T_F m_i}/(eB)$.
The gradient length scale enters as $g_n= L_{\perp} / L_n$. 
Two possibilities for length scale normalization are appropriate, setting
$L_{\perp}$ either to $\rho$ or to $L_n$. 

The parallel derivative acting on the current is not explicitly evaluated
here, but rather determined by a fixed parallel wavenumber $k_{||}$ through a
dissipative coupling parameter $d= (\omega_{ce}/\nu_{ei} \delta_0)k_{||}^2$. 
Accordingly, we can rewrite eqs.~(\ref{e-dens2}) and (\ref{div-j}) 
as a set of normalized quasi-two-dimensional equations 
\begin{eqnarray}
\partial_t \Omega + [ \phi, \Omega ] &=&
d ( \phi - \Lambda n )  - \kappa (\Lambda^{\!\ast} n) \label{qhw1} \\
\partial_t n + [ \phi, n ] &=& 
d ( \phi - \Lambda n )  - g_n \partial_y \phi 
 + \kappa ( \phi - \Lambda^{\! \ast} n )
 \label{qhw2} 
\end{eqnarray}
where $\Omega = \nabla_\perp^2 \phi$ and $[a,b] =\hat {\bf z} \times \nabla a \cdot \nabla b$
represent the ion vorticity and Poisson's bracket, respectively. 
The quantum diffraction effect enters through 
$\Lambda = 1 - \beta^2\nabla^2$ and $\Lambda^{\! \ast} = 1 + \beta^2\nabla^2$,
with $\beta = \lambda_q/\rho$. It acts as a finite de Broglie length
screening effect on density fluctuations, similarly to the established FLR
gyro screening and Debye screening effects.
Eqs.~(\ref{qhw1}) and (\ref{qhw2}) are the semi-classical dense plasma
generalization of the Hasegawa-Wakatani equations \cite{Hasegawa83}.

\begin{figure}
\includegraphics[width=8.5cm]{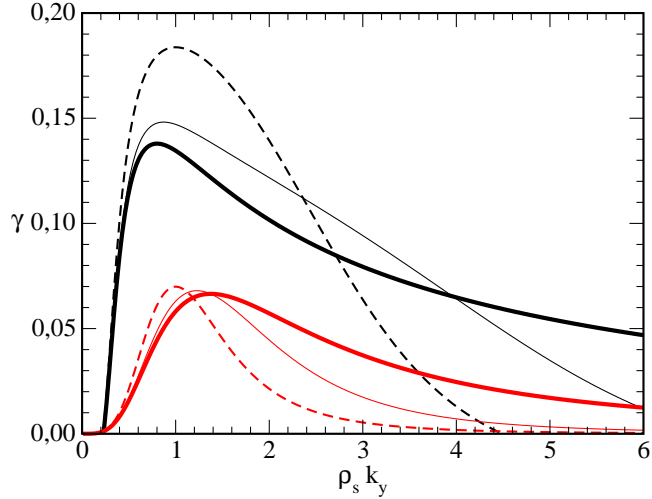}
\caption{\label{fig-lin} \sl The linear growth rate  $\gamma (\rho k_y)$ for $d=0.1$
  (black) and $d=2$ (red), with the de Broglie factors $\beta=0$ (bold curves),
  $\beta=0.2$ (thin curves) and $\beta=1.0$ (dashed curves).}
\end{figure}

First, we derive the linear dispersion relation from Eqs.  (\ref{qhw1}) and (\ref{qhw2}) for $\kappa=0$. 
For this purpose, we neglect nonlinear terms and suppose that $\phi$ and $\tilde{n}$ are proportional to
$\exp(-i\omega t + i {\bf k} \cdot {\bf r})$, where $\omega$ and ${\bf k}$ are the 
normalized frequency and wave vector, respectively. We have 
\begin{eqnarray}
k^2 \omega^2 +id (1+ \Lambda k^2) \omega  - id \Lambda g_n k_y = 0 
\label{dispersion}
\end{eqnarray}
where $\Lambda = (1+\beta^2  k^2)$. 
In the adiabatic limit $d \gg 1$ we obtain $\omega_{\ast} = \Lambda g_n k_y / (1+ \Lambda k^2)$
as the electron diamagnetic drift frequency including quantum corrections.
Inserting $\omega = \omega_R + i \gamma_0$ into the dispersion relation and
solving for the imaginary component in the limit $\gamma_0 \ll \omega_{\ast}$ 
we obtain the linear growth rate for weakly nonadiabatic quantum drift waves:
\begin{equation}
\gamma_0 = { g_n^2 k_y^2 \Lambda^2 k^2 \over d (1+\Lambda k^2)^3 } \equiv
       {\omega_{\ast}^2 \over 2 \omega_0}
\label{growth}
\end{equation}
with $\omega_0 = d(1+\Lambda k^2)/(2k^2)$.
The exact solution 
\begin{equation}
\gamma = \omega_0 \left[ 1 \pm {\sqrt{2} \over 2} 
         \sqrt{ 1 + \sqrt{ 1+ 16(\gamma_0/\omega_{\ast})^2 } } \right]
\end{equation}
for the imaginary part of the frequency from dispersion relation
(\ref{dispersion}) is shown  in Fig.~\ref{fig-lin} for $k_x=0$, $g_n=1$,
$d=0.1$ (black) or $d=2$ (red), and various values of $\beta$.   
The maximum growth is found around $\rho k_y \sim 1$.
In the more strongly collisional case ($d=0.1$), the maximum growth rate is
increased by around one third for $\beta=1$ compared to $\beta=0$ and is
shifted towards slightly smaller wavelengths.  
For lower collisionality ($d=2$), the growth rate remains nearly constant when
$\beta$ is raised,  but the maximum shifts to slightly lower $k_y$, while the
de Broglie screening significantly reduces the growth rates for $\rho k_y > 1$.  
For $d>2$ the growth rate $\gamma \approx \gamma_0$ is well approximated by
the weakly adiabatic solution.

Next, we numerically solve the nonlinear quantum drift wave Eqs. (\ref{qhw1}) and (\ref{qhw2}) 
for a uniform magnetic field ($\kappa=0$). For time stepping, an explicit third order Karniadakis scheme
\cite{Karniadakis} is applied, and the Poisson bracket $[a,b] = (\partial_x a)(\partial_y b)  
- (\partial_y a)(\partial_x b)$ is evaluated with the energy and enstrophy conserving Arakawa 
method \cite{Arakawa}. The numerical method is equivalent to the one
introduced in Ref.~\cite{Naulin03}.   
A hyper-viscous operator, e.g. associated with strong ion coupling effects, $\nu^4\nabla^4$ with 
$\nu^4= -10^{-4}$ is added  for numerical stability into the right-hand side of both 
Eqs. \ref{qhw1} and \ref{qhw2}, acting on $\Omega$ and $n$, respectively. 
The Poisson equation is solved spectrally.  The equations are for the following computations 
discretized on a doubly periodic 1024 $\times$ 1024 grid with box-dimension $L = 64 \rho$. 
Spatial scales are in units of $\rho$.  
Here we choose $L_{\perp} = L_n$ and thereby gain fluctuations $\phi$ and $n$
in the order of unity (for $\delta_o \ll 1$ and $ e \phi / T_F \ll 1$, $n/n_0
\ll1$)) and accordingly set $g_n\equiv 1$. The nominal parallel coupling
parameter is set to $d=0.5$.  
The computations are initialized with a Gaussian density perturbation which
transiently develops into a drift wave during a quasilinear instability phase
and finally saturates nonlinearly into a fully developed turbulent state.

\begin{figure}
\includegraphics[width=8.5cm]{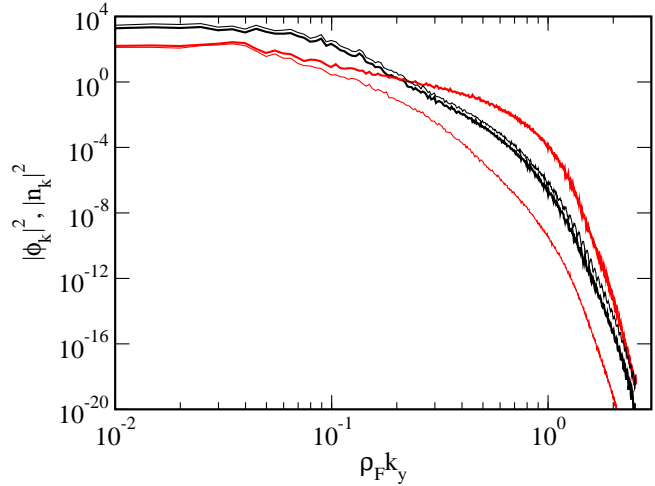}
\caption{\label{fig-specs} \sl Spectra $|\phi(k)|^2$ (black) and $|n(k)|^2$ (red) of
 a fully developed quantum drift wave turbulence for $\beta=0$ (bold) and $0.5$
  (thin).} 
\end{figure}

In Fig.~\ref{fig-specs} the wavenumber spectra of the electrostatic potential
fluctuations $|\phi(k)|$ (black) and density fluctuations $|n(k)|$ (red)in the fully developed
turbulent state are shown for $\beta=0$ (bold) and 0.5 (thin). The spectral
properties of $|\phi(k)|$ remain similar with larger $\beta$, while its fluctuation level is
increased for finite $\beta$ for all wavelengths as a consequence of the
larger growth rates. The density $|n(k)|$ on the other hand shows a strongly
reduced level for all but the largest scales.
For a larger dissipative coupling parameter $d$ (corresponding to a lower
electron-ion collision frequency) the electrostatic potential fluctuations are
more strongly coupled to the density perturbations and experience similar screening.

\begin{figure}
\includegraphics[width=8.5cm,height=5.0cm]{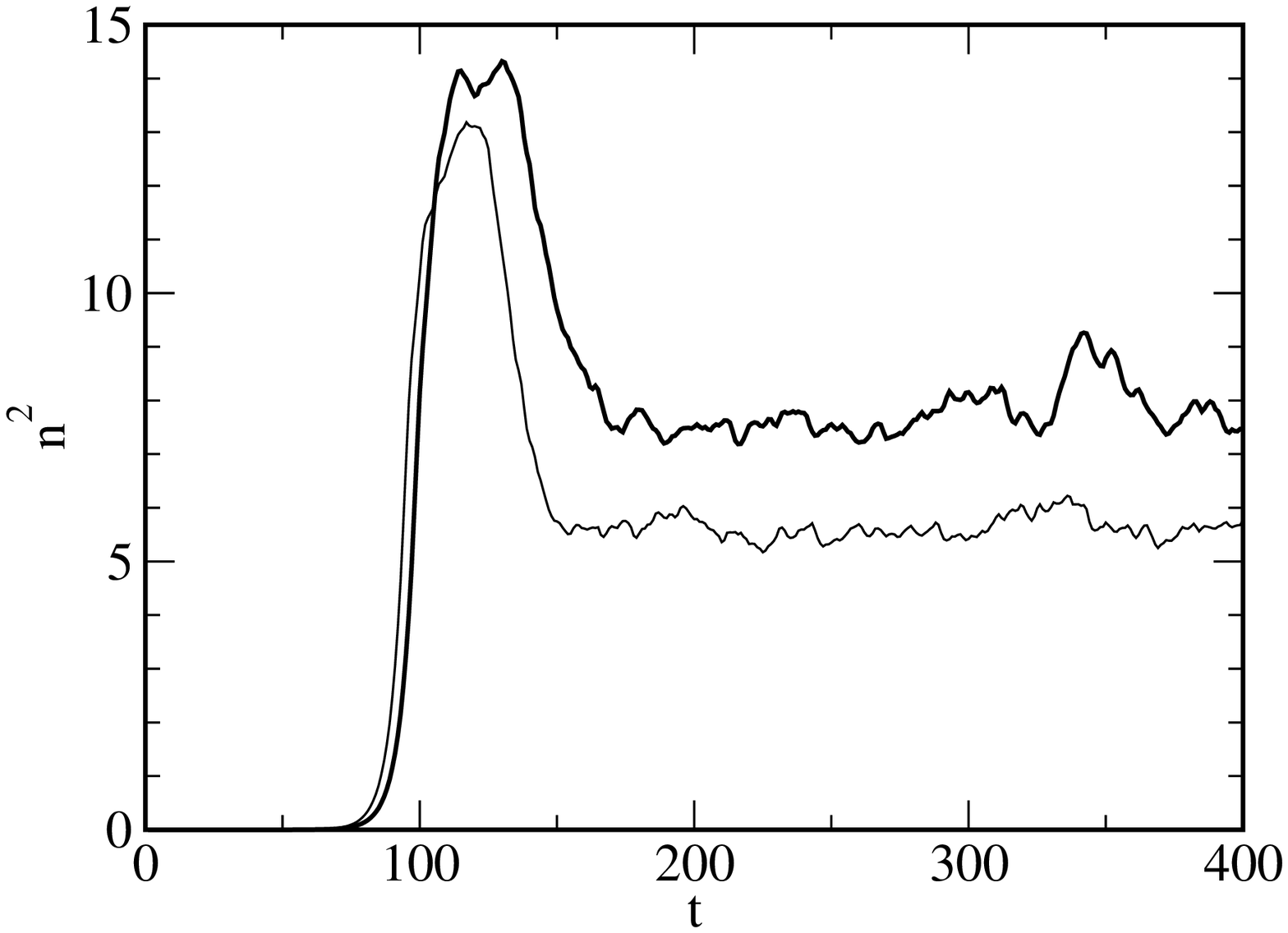}\\
\includegraphics[width=8.5cm,height=5.0cm]{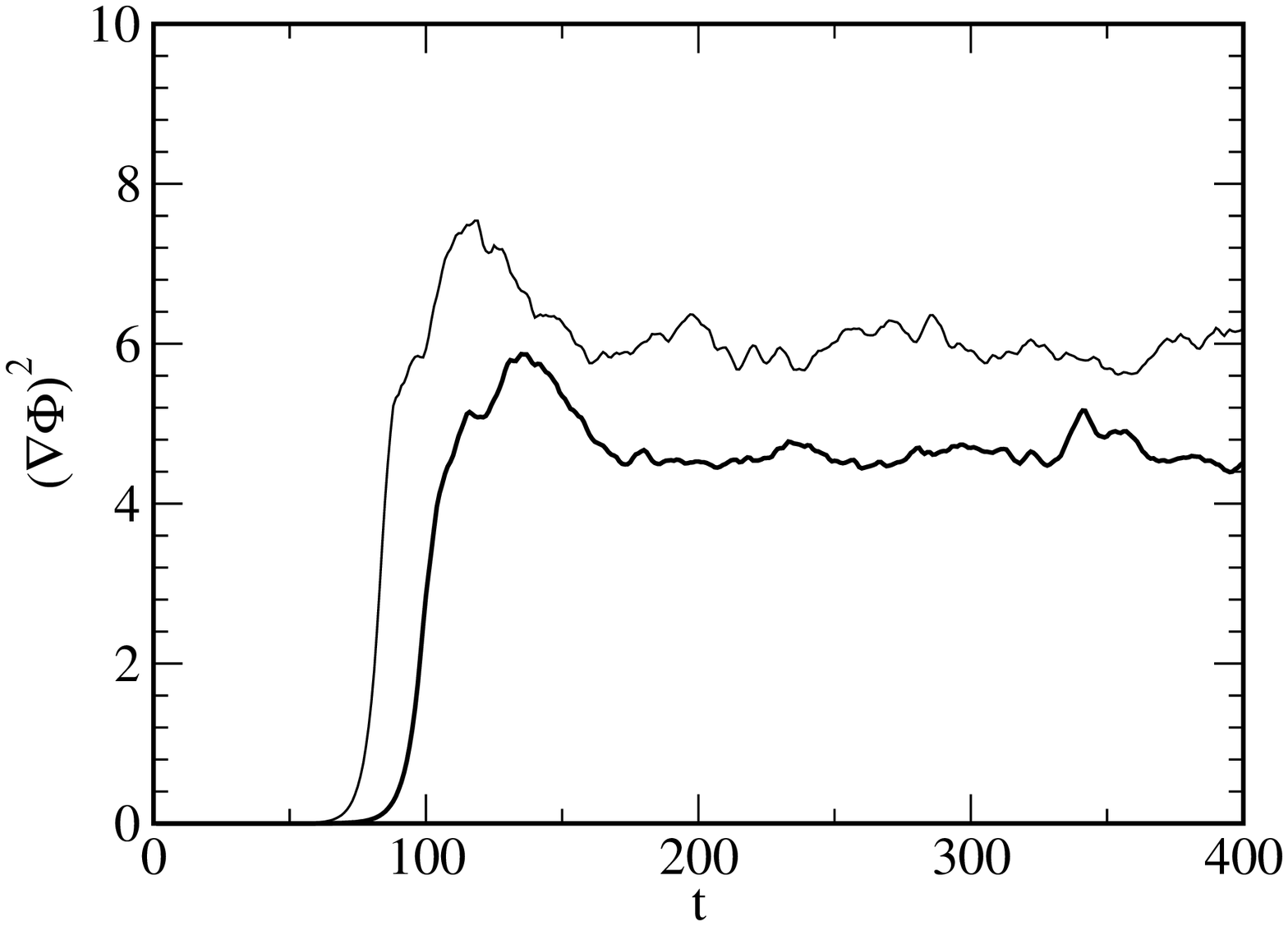}
\caption{\label{fig-traces} \sl
Time traces of the potential energy $n^2$ (top) and the kinetic energy $(\nabla
\phi)^2$ (bottom) for $\beta=0$ (bold) and 0.5 (thin)} 
\end{figure}

\begin{figure}
\includegraphics[width=5.0cm]{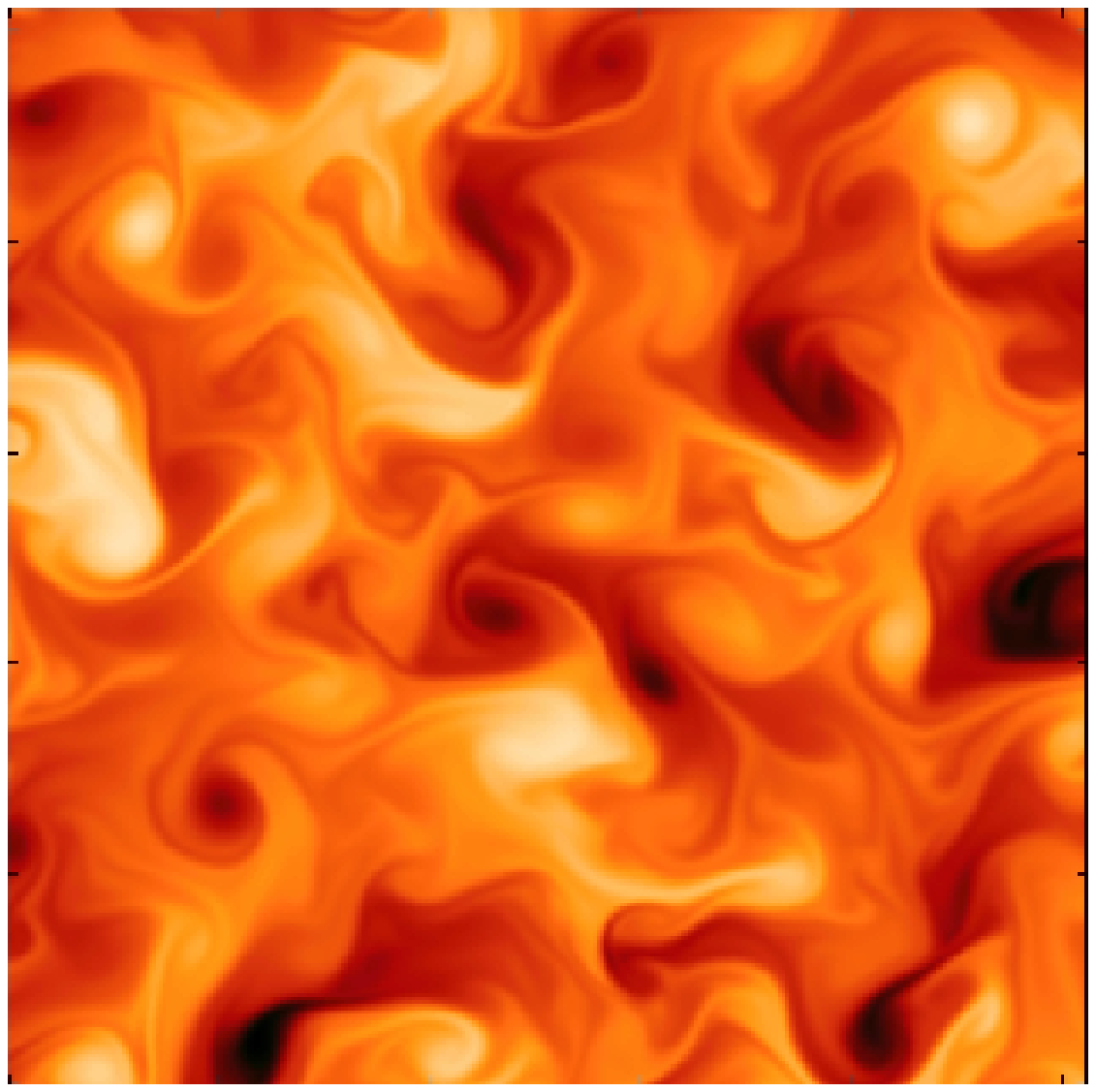}
\includegraphics[width=5.0cm]{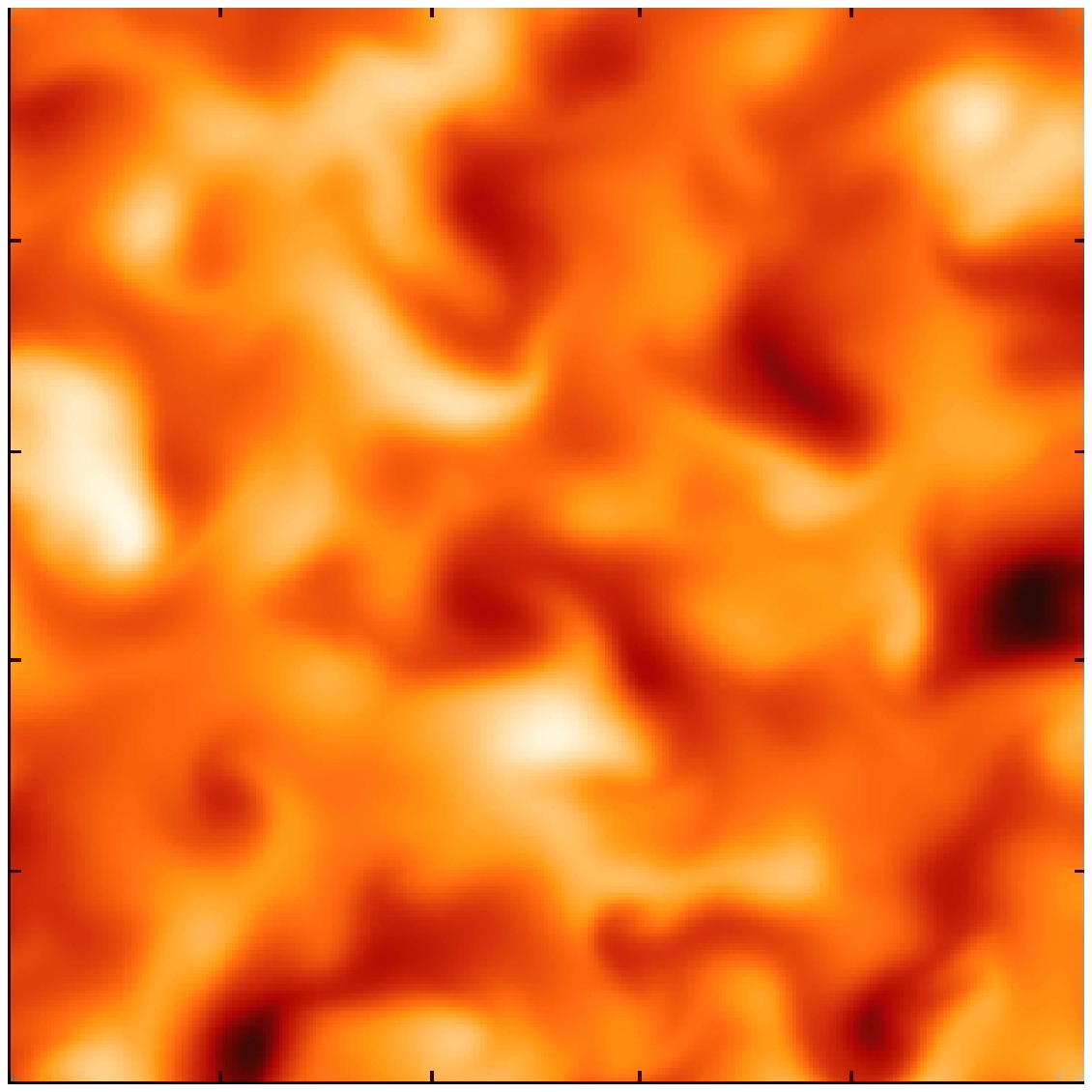}\\
\includegraphics[width=5.0cm]{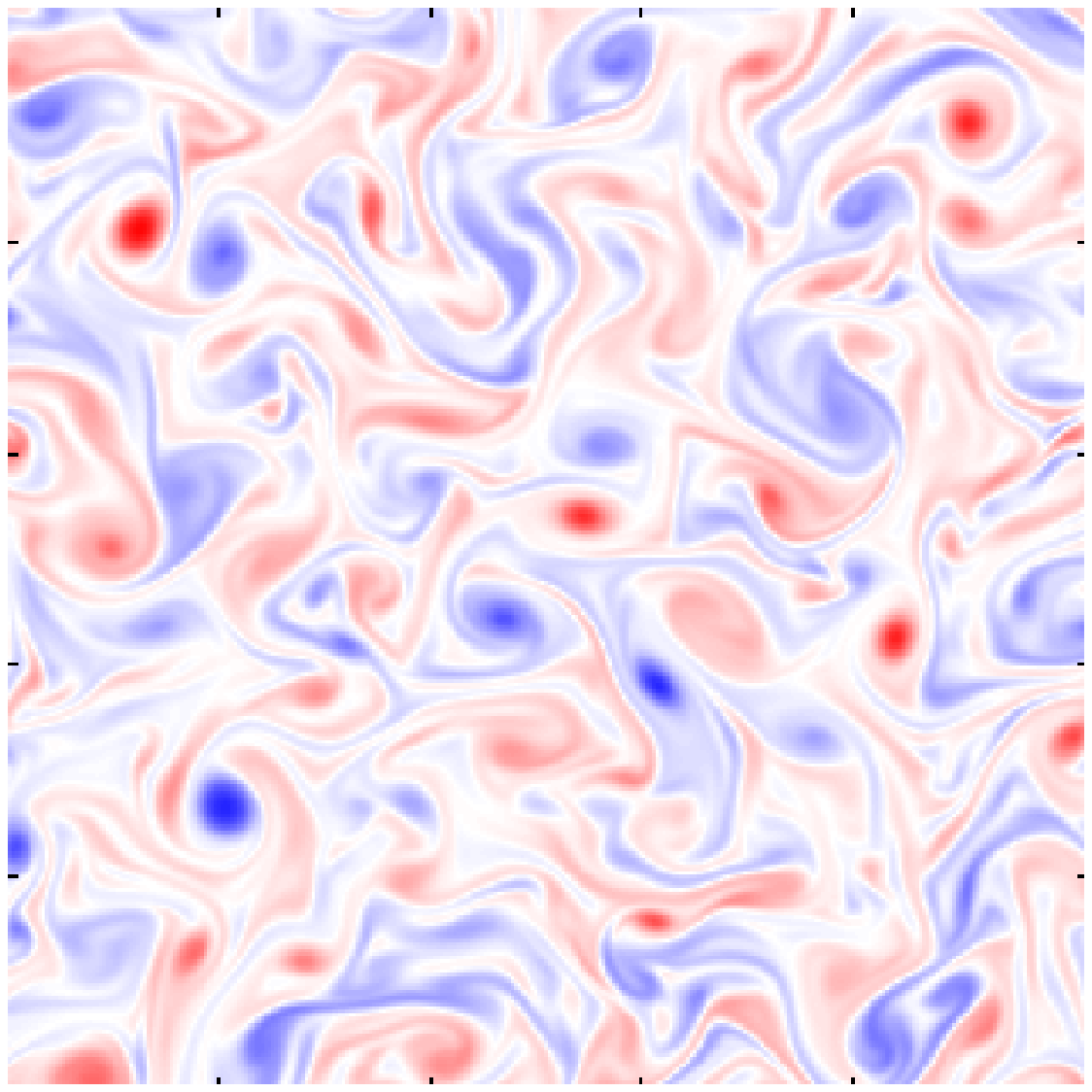}
\includegraphics[width=5.0cm]{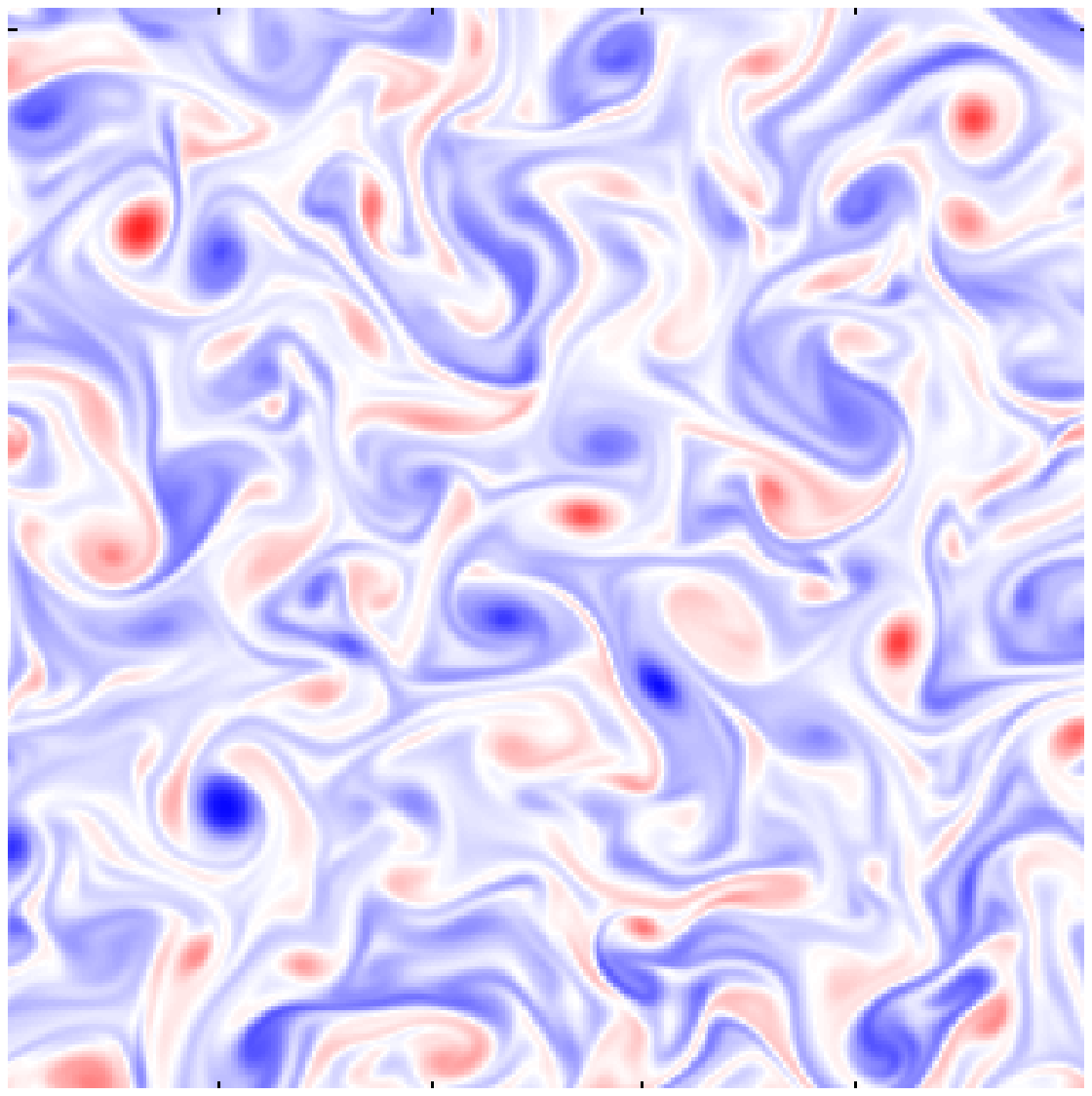}
\caption{\label{fig-plots} \sl 
2-D plots of the density $n(x,y)$ (top) and the ion vorticity $\Omega(x,y)$ (bottom) for
$\beta=0$ (left) and $\beta=0.5$ (right). Only a quarter section ($512 \times
512$) of the computational domain is shown.} 
\end{figure}

Our simulations show for the first time the effect of de Broglie screening on
dense plasma turbulence. 
This novel quantum diffractive screening mechanism also affects the turbulent fluctuation energies.
The total energy $E = (1/2) \int dV \left( n^2 + (\nabla \phi)^2 \right)$
is only slightly changed (by around $\pm 10\%$, depending on the plasma parameters) for
$\beta \leq 0.5$ compared to $\beta=0$. Its components, the potential energy $E_n 
= (1/2) \int dV n^2$ and the kinetic energy $E_k = (1/2) \int dV (\nabla
\phi)^2 )$, are however strongly and reciprocally influenced.
Figure ~\ref{fig-traces} shows $E_n(t)$ (top) and $E_k(t)$ (bottom) for $\beta=0$
(bold) and 0.5 (thin). While the density perturbations decrease
with increasing $\beta$, the velocity fluctuations increase. 
The quantum effect on the drift wave turbulence is visualised in 2-D plots of
the turbulent fluctuations in Fig.~\ref{fig-plots} for $\beta=0$ (left column)
and 0.5 (right column): the de Broglie screening affects density perturbations
(top row), but conserves the fine structure in the ion vorticity (bottom row). 

In conclusion, we have derived and solved nonlinear DW equations for 
nonuniform Fermi magnetoplasmas that are collisional. The present nonlinear equations 
include the new physics of quantum forces involving the pressure of degenerate
electron fluids and overlapping electron wavefunctions over nanoscales.
The quantum regime is characterized by $\beta = \lambda_q/\rho$, which is the 
ratio between the de Broglie length and the drift scale at the Fermi electron temperature. 
The quantum effect acts as a de Broglie screening on density perturbations for small 
wavelengths, and increases the growth rate and turbulent velocity fluctuations
across the whole spectrum. The DW vortex structures occur at nanoscales, which are of the 
the order of several Fermi sound gyroradius $\rho$. 
The novel de Broglie effect on turbulent drift vortices is of a similar quality as
finite Larmor radius (FLR) corrections in gyrofluid or gyrokinetic simulations of
warm ion (ITG) drift wave turbulence \cite{Dorland93}, and as Debye shielding effects on
electron gyro scale (ETG) turbulence \cite{Jenko02}.

In closing, the results of the present investigation were primarily motivated by
theoretical interest, but may be useful in understanding turbulence and
coherent structures that arise in nonuniform dense magnetoplasmas, such as
those in the atmospheres of white dwarfs \cite{Shapiro83,Koester90,Lai01,Hansen04}. 
Fully developed DW turbulence can regulate the cross-field plasma particle
transport over nanoscales. 

For example, a mixing length estimate of DW turbulent convective particle diffusivity $D_{turb}
\approx \omega_{\ast} \rho(T_F)^2$ in comparison with collisional diffusive cross-field transport
$D_{coll} \approx \nu_{ei} \rho_i(T_i)^2$ in a strongly magnetized plasma
depends essentially on the the (local or radially global) density gradient
length, which is subject to a large uncertainty. 
Assuming a white dwarf surface density $n \sim 10^{30} cm^{-3}$, the collisionality is in
the order of $\nu_{ei} \sim 10^{11} s^{-1}$ (independent of density in a
degenerate plasma). The Fermi sound speed is in the order of $c_s \sim 10^7
cm/s$ (smaller than 0.1~$c$, justifying the present non-relativistic
approximation), and the radially global density gradient length is in the
order of $L_n^{-1} \sim 10^4 cm^{-1}$, giving $\omega_{\ast} = c_s / L_n \sim
10^3 s^{-1}$, so that $\nu_{ei} / \omega_{\ast} \sim 10^6$. On the other hand,
$\rho(T_F)^2 / \rho_i(T_i)^2 = T_F / T_i \sim 10^6$ is of the same ratio
(for an ion temperature $T_i$ in the order of $10^4 K$), so that $D_{turb}
\approx D_{coll}$ are in the same range. 
Passive cross-field advection of heavier trace ions by small-scale DW
turbulence could contribute to counter sedimentation in cool white dwarfs,
even in the absence of thermal convection or radiative acceleration.


\section*{Acknowledgements}
This work was supported by the Austrian Science Fund (FWF) project no.~Y398,
 and by a junior research grant from University of Innsbruck.

\end{document}